\documentclass[preprint,showpacs,preprintnumbers,superscriptaddress,nofootinbib]{revtex4}
\usepackage{bm,amsmath,amssymb}
\usepackage{graphicx,subfigure}
\usepackage[usenames,dvipsnames]{color}
\usepackage{soul}
\setulcolor{Red}
\setstcolor{Red}
\sethlcolor{Yellow}

\newcommand{\beq}{\begin{equation}}
\newcommand{\eeq}{\end{equation}}
\newcommand{\be}{\begin{equation}}
\newcommand{\ee}{\end{equation}}
\newcommand{\ber}{\begin{eqnarray}}
\newcommand{\eer}{\end{eqnarray}}
\newcommand{\berr}{\begin{eqnarray*}}
\newcommand{\eerr}{\end{eqnarray*}}
\newcommand{\ba}{\begin{array}}
\newcommand{\ea}{\end{array}}

\newcommand{\old}[1]{}

\begin{document}

\vspace*{1cm}
\title{
Signatures of charmonium modification in spatial correlation functions
}

\author{F.\ Karsch}
\email[E-mail: ]{karsch@bnl.gov}
\affiliation{Fakult\"at f\"ur Physik, Universit\"at Bielefeld, Bielefeld 33615, Germany} 
\affiliation{Brookhaven National Laboratory, Upton NY 11973, USA}
\author{E.\ Laermann} 
\email[E-mail: ]{edwin@physik.uni-bielefeld.de}
\affiliation{Fakult\"at f\"ur Physik, Universit\"at Bielefeld, Bielefeld 33615, Germany} 
\author{Swagato\ Mukherjee}
\email[E-mail: ]{swagato@bnl.gov}
\affiliation{Brookhaven National Laboratory, Upton NY 11973, USA}
\author{P.\ Petreczky}
\email[E-mail: ]{petreczk@bnl.gov}
\affiliation{Brookhaven National Laboratory, Upton NY 11973, USA}

\begin{abstract}
We study spatial correlation functions of charmonium in 2+1 flavor QCD using an
improved staggered formulation. Contrary to the temporal correlation functions the
spatial correlation functions exhibit a strong temperature dependence above the QCD
transition temperature. Above this temperature they are sensitive to temporal
boundary conditions. Both features become significant at a temperature close to
$1.5 T_c$ and suggest corresponding modifications of charmonium spectral functions.
\end{abstract}

\preprint{BNL-97047-2012-JA}
\preprint{BI-TP 2012/09}

\maketitle

\section{Introduction}

Quarkonium suppression was proposed long ago by Matsui and Satz as a signal for
deconfinement in heavy ion collisions \cite{MS86}. The basic idea behind this
proposal was the fact that at sufficiently   high temperatures color screening will
prevent formation of heavy quark bound states (for recent reviews on this topic see
\cite{qwg,qgp4}).  Charmonium suppression was indeed observed in heavy ion
experiments at SPS \cite{sps} and at RHIC \cite{rhic}, and is now being intensively
studied also at the LHC \cite{lhc}.  In order to interpret current and future
experimental findings it is very important to know (among other things) the in-medium
properties of heavy quark anti-quark pairs.

The notion of bound state melting can be rigorously formulated in terms of the
spectral functions. Charmonium dissociation corresponds to gradual broadening and
eventual disappearance of the bound state peaks in the corresponding meson spectral
functions. Spectral functions are related to the Euclidean time meson correlation
functions and can be studied using lattice
QCD calculations. The standard approach to obtain information about the spectral
functions  from calculations of temporal  correlators relies on the Maximum Entropy
Method (MEM) \cite{asakawa01,ineslat01}.  Based on this approach, analyses in
quenched QCD led to the conclusion that 1S charmonium states may survive up to
temperatures as high as $1.6T_c$, with $T_c$ being the deconfinement temperature of
the SU(3) gauge theory \cite{umeda02,asakawa03,datta04,jako06} (for a similar
analysis in 2 flavor QCD see \cite{alton07}). Other lattice QCD studies based on
different techniques, e.g.  variational analysis, led to similar conclusions
\cite{Iida:2006mv,Ohno:2011zc}. The conclusion that charmonium states survive in a
certain temperature range above the QCD transition temperature is largely based on
the observation that the corresponding temporal meson correlation functions show a
weak temperature dependence across the transition
\cite{datta04,jako06,umeda07,petr08,ding10}.  Moreover, at high temperatures in
particular the vector spectral function receives a contribution near zero frequency,
$\omega \simeq 0$, which corresponds to the transport of heavy quarks.  This
near-zero-mode of the spectral function gives rise to a contribution to the temporal
correlation function that is (almost) constant in Euclidean time and is responsible
for most of the temperature dependence of temporal meson correlators
\cite{umeda07,petr08}. Thus, the temperature dependent effects due
to in-medium modifications and dissolution of charmonium states appear to be small in
temporal correlation functions \footnote{Using NRQCD at finite temperature some
evidence for melting of bottomonium P-states, that have same size as the ground state
charmonium, was obtained \cite{Aarts:2010ek}.}

Contrary to the lattice analysis of temporal meson correlators, potential model
studies that use the static quark anti-quark correlators calculated in lattice QCD
\cite{f1} as an input into the Schr\"odinger equation, predict melting of charmonium
bound states at temperatures slightly above the QCD transition temperature
\cite{digal01,blaschke04,mocsy07}.  Furthermore, analyses of quarkonia at non-zero
temperature within the effective field theory approach revealed that in addition to
the modification of  its real part, the potential also acquires an imaginary part
\cite{laine06,nora08}. The imaginary part plays an important role in quarkonium
dissociation.  Even when one takes into account the uncertainties in relating the
imaginary part of the potential to the correlation functions of a static quark
anti-quark pair calculated in lattice QCD, a non-vanishing imaginary part leads to
the dissolution of at least the charmonium bound states \cite{miao10}. A similar
conclusion was reached in the analysis of the charmonium spectral functions using the
T-matrix approach \cite{riek}. Remarkably though, it was shown in the framework of
potential models that the melting of quarkonium states does not result in large
changes of the temporal correlators and the correlators seemed to be temperature
independent to very good approximation \cite{mocsy07}.  Thus both potential models
and direct lattice QCD calculations suggest only a weak temperature dependence of the
temporal meson correlators across the transition. 

In order to extract detailed features of the spectral functions, {\sl e.g.}
broadening of bound state peaks or melting of states, from such weakly temperature
dependent temporal correlators one needs high precision data at a large number of
Euclidean time separations. The most recent high statistics (quenched) lattice
calculation of charmonium spectral functions using MEM reached lattice spacings down
to $a = 0.01$ fm \cite{ding10}. This study found no evidence for bound state
peaks at $T\simeq 1.46T_c$, suggesting charmonium melting to take place at a
temperature somewhat smaller compared to the earlier MEM based lattice QCD analyses.
Nevertheless, it is desirable to have alternative observables that may
provide additional information on the in-medium properties of heavy quark bound
states.

In this paper we therefore analyze spatial charmonium correlators at non-zero
temperature.  Following \cite{boyd94} we also study the dependence of the spatial
meson correlation functions on the temporal boundary condition.  We have found that
the behavior of spatial charmonium correlators and their dependence on the boundary
condition is consistent with significant modifications of charmonium states at
sufficiently high temperatures. Some preliminary results of this work were presented
previously in \cite{Muk1,Muk2}. We also note that some exploratory studies of spatial
charmonium correlators in quenched QCD were reported in \cite{datta04}. Our findings
in 2+1 flavor QCD are qualitatively consistent with these calculations.

\section{Spatial correlation functions }
 
Correlation functions of a meson operator $J=\bar q \Gamma q$ along the spatial
direction $z$ are defined as
\begin{equation}
G(z,T)=\int dx dy \int_0^{1/T} d \tau \langle J(x,y,z,\tau) J(0,0,0,0) \rangle .
\end{equation}
The matrix $\Gamma$ is a product of Dirac matrices and fixes the quantum number of
the meson. The spatial correlation function is related to the meson spectral function
at non-zero spatial momentum
\begin{equation}
G(z,T)=\int_{0}^{\infty} \frac{2 d \omega}{\omega} \int_{-\infty}^{\infty} d p_z e^{i p_z z} \sigma(\omega,p_z,T)
\end{equation}
Thus the temperature dependence of the spatial correlation function also provides
information about the temperature dependence of the spectral function.  Medium
effects are expected to be the largest at distances that are larger than $1/T$.  At
these distances $G(z,T)$ decays exponentially and this exponential decay is governed
by a screening mass $M_{scr}$. It is also pertinent to note here that the above
relation suggests, unlike the temporal correlation functions, an
$\omega\delta(\omega)$ type zero-mode contribution to the spectral function does not lead
to a non-decaying constant contribution to the spatial correlation functions and only
leads to a contact term.

If there is a lowest lying meson state of mass $M$, this is signaled by a peak in the
spectral function, $\sigma(\omega,p_z,T) \sim \delta(\omega^2-p_z^2-M^2)$. This bound
state peak in the spectral function determines the long distance behavior of the
spatial meson correlation function. We thus have $M_{scr}=M$. On the other hand, at
very high temperatures the charm quark and anti-quark are not bound and the meson
screening mass is given by $2 \sqrt{(\pi T)^2+m_c^2}$, where $m_c$ is the charm quark
mass and $\pi T$ is the lowest fermionic Matsubara frequency.  The above value of the
screening mass is a consequence of the anti-periodicity of the fermion fields along
the time direction and forces the charm quark (anti-quark) to pick up at least a $\pi
T$ contribution from the lowest non-vanishing Matsubara frequency at non-zero
temperatures.  The transition between these two limiting behaviors of the screening
masses may serve as an indicator for significant modifications and ultimately
dissolution of the meson states.  

Furthermore, if the $c \bar c$ pair forms a mesonic bound state the screening mass is
not expected to be sensitive to the anti-periodic temporal boundary condition due to
the bosonic nature of the basic degrees of freedom.  In this case using a periodic
temporal boundary condition for the fermions is expected to give the same screening
mass as for the case of an anti-periodic temporal boundary condition. Since for
the non-interacting theory with a periodic temporal boundary condition
$M_{scr}=2m_c$, a comparison with the usual anti-periodic case of
$M_{scr}=2\sqrt{(\pi T)^2+m_c^2}$ will also facilitate in the identification of free
theory like behavior of the spatial correlation functions. Thus studying the
dependence of the screening masses on the temporal boundary conditions may also
provide some additional information about the existence of bound states.
\begin{figure}
\begin{center}
\includegraphics[scale=0.75]{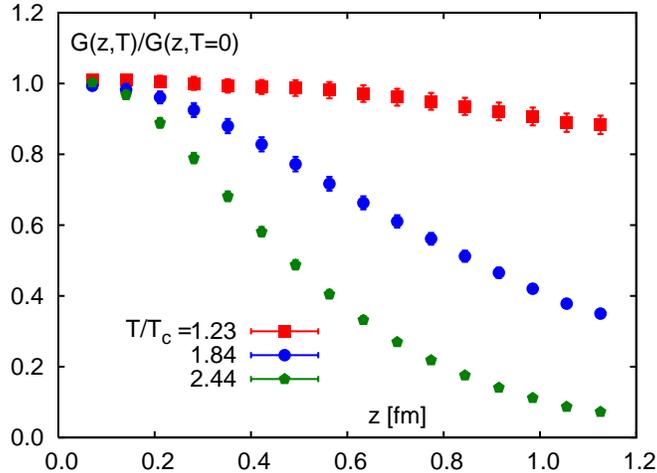} 
\caption{Ratio of the spatial pseudo-scalar correlators to the corresponding zero
temperature correlators calculated for $a^{-1}= 2.8$~GeV at three different
temperatures.  Calculations have been performed on lattices with temporal extent
$N_\tau=6,~8$, and $12$. In units of the transition temperature, these correspond to
$T/T_c= 1.23,~1.84$, and $2.44$.} 
\label{fig:corr}
\end{center}
\end{figure}

\section{Numerical results}

We performed numerical calculations in 2+1 flavor QCD using the improved staggered p4
action \cite{p4-1,p4} on $32^3 \times N_{\tau}$ lattices with $N_{\tau}=6, 8 ,12$ and
$32$. Some calculations have also been performed on $24^3 \times 6$ lattice.  We
mostly used the gauge field configurations generated for the study of the QCD
equation of state \cite{rbcbi07,hoteos}. The strange quark mass $m_s$ was fixed to
its physical value, while for the light quark masses we used $m_s/10$ corresponding
in the continuum limit to a pion mass of about $220$~MeV.  The lattice spacing was
fixed using the value $r_0=0.469$~fm \cite{gray05} for the Sommer scale
\cite{sommer}. A detailed discussion of the choice of the lattice parameters is given
in Refs. \cite{rbcbi07,hoteos}.  Since for lattice spacings used in this study cutoff
effects may be still significant we will present our results in terms of the reduced
temperature $T/T_c$.  For the transition temperature for $N_{\tau}=6$ and $8$
lattices we will use the values $T_c=198$~MeV and $T_c=191$~MeV from Ref.
\cite{rbcbi06,hotqcd2}.  From the $O(N)$ scaling analysis presented in \cite{hotqcd2}
we estimate $T_c \simeq 160$~MeV in the continuum limit for light quark masses of
$m_s/10$. Assuming a $1/N_{\tau}^2$ dependence of $T_c$ for $N_{\tau}>8$ we estimate
that $T_c\simeq 174$~MeV for $N_{\tau}=12$.  We also used the staggered p4 action for
the valence charm sector. The valence charm quark masses were fixed to reproduce the
physical value of the $J/\psi$ mass at $T=0$ \cite{cheng-1,cheng-2}. 

The staggered fermion formulation describes four valence quark flavors in the
continuum limit. Meson operators in this formalism are written as $J=\bar q (\Gamma^D
\times \Gamma^F) q$, where $\Gamma^D$ and $\Gamma^F$ are products of the Dirac
matrices and describe the spin flavor structure of the meson \cite{kluberg}.  Here we
consider only local meson operators for which $\Gamma^D=\Gamma^F=\Gamma$, since
these are the cheapest to calculate. As in most lattice QCD studies only the
quark-line connected part of the correlation function is taken into account; however,
for heavy quarks the effect due to the disconnected part is expected to be small. The
meson operators can be written in terms of staggered quark fields, $\chi$ and $\bar
\chi$ as $J= \tilde \phi(X) \bar \chi(X) \chi(X),~ X=(x,y,z,\tau)$
\cite{golterman,altmeyer}.  Different meson channels/operators are fixed by
appropriate choice of the phase factor $\tilde \phi(X)$.  Each choice of $\tilde
\phi(X)$ corresponds to a pair of $\Gamma$ matrices related to positive and negative
parity states that show up as oscillatory and non-oscillatory terms in the
correlation function \cite{altmeyer}. In our analysis we used both point and wall
sources.

In this paper we mainly discuss the simplest case corresponding to $\tilde
\phi(X)=-(-1)^{\tau+x+y}$.  We will comment on other choices of $\tilde \phi(X)$
later.  The correlator corresponding to $\tilde \phi(X)=-(-1)^{\tau+x+y}$ is
dominated only by pseudo-scalar meson states (the oscillatory contribution vanishes)
and thus is the easiest to analyze.  For this reason we refer to the corresponding
correlator as the pseudo-scalar correlator.  In the zero temperature limit this
correlator corresponds to the spin singlet S-wave charmonium state $\eta_c$. First we
are interested in the change of the correlation functions in the high temperature
regime of QCD relative to the zero temperature correlator.  In Fig.~\ref{fig:corr} we
show the ratio of the spatial correlators calculated for three different temperatures
corresponding to $N_{\tau}=6,~8$ and $12$ and for fixed lattice spacing
$a^{-1}=2.8$~GeV. As one can see from the figure the correlator shows significant
change already at a temperature of  $T/T_c\simeq 1.2$.  As the temperature increases
further the deviations from the zero temperature correlator become more prominent.
This is different from the case of the temporal correlators, where very little change
is seen even at the highest temperature. This is partly due to the fact that larger
separations can be probed in the spatial direction. Indeed, for separations smaller
than $1/(2T)$, available  in the case of temporal correlators, the temperature
dependence is very moderate.

We have studied the long distance behavior of the spatial correlators and using simple 
exponential fits  extracted
the corresponding screening masses as well as the amplitudes.  In the zero
temperature limit the latter are proportional to the square of the wave functions at
the origin. At infinite temperature, i.e. in the free case limit, the leading
behavior of the correlation function has the form
$\sim e^{-m_{scr} z}/z$ \cite{florkowski}. However, a $1/z$ dependence of the amplitude
or, corrolary, a decrease of the effective screening mass
$m_{scr}(z)=\ln{G(z)/G(z+1)}$ proportional to $\ln (1+1/z) \simeq 1/z$ has not been observed.

\begin{figure}
\begin{center}
\subfigure[]{ \includegraphics[scale=0.65]{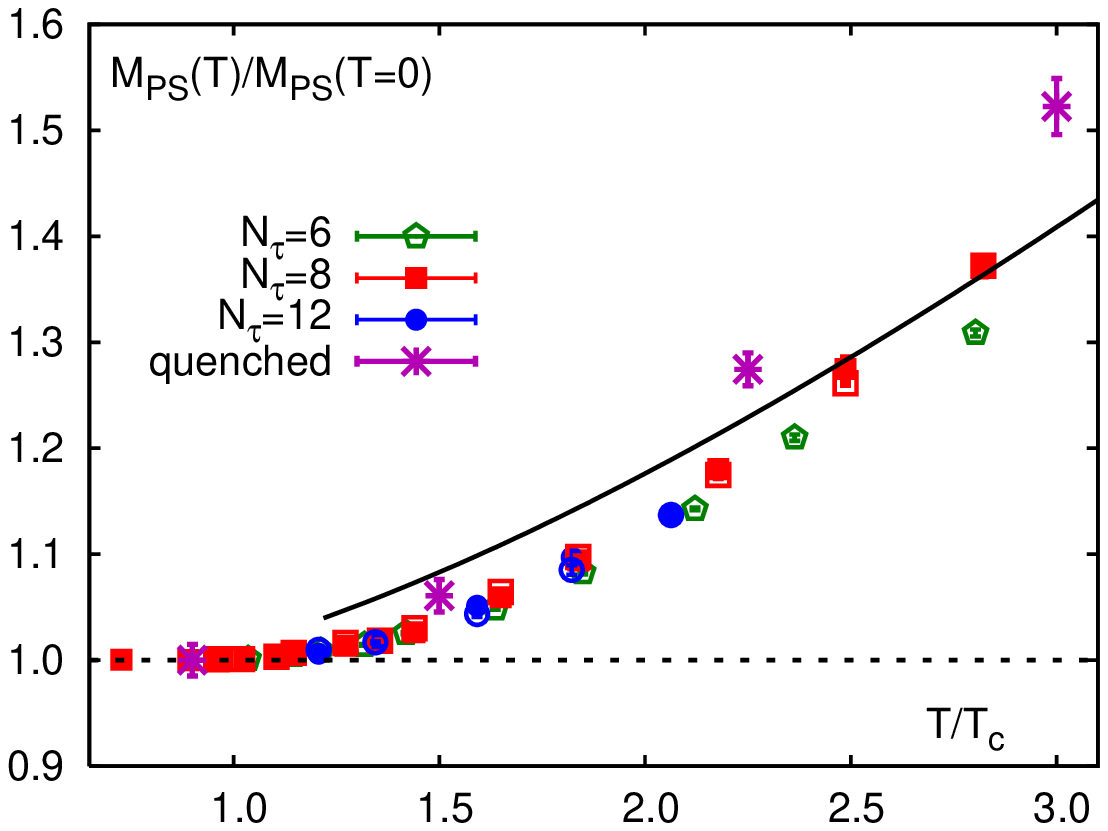} }
\subfigure[]{ \includegraphics[scale=0.65]{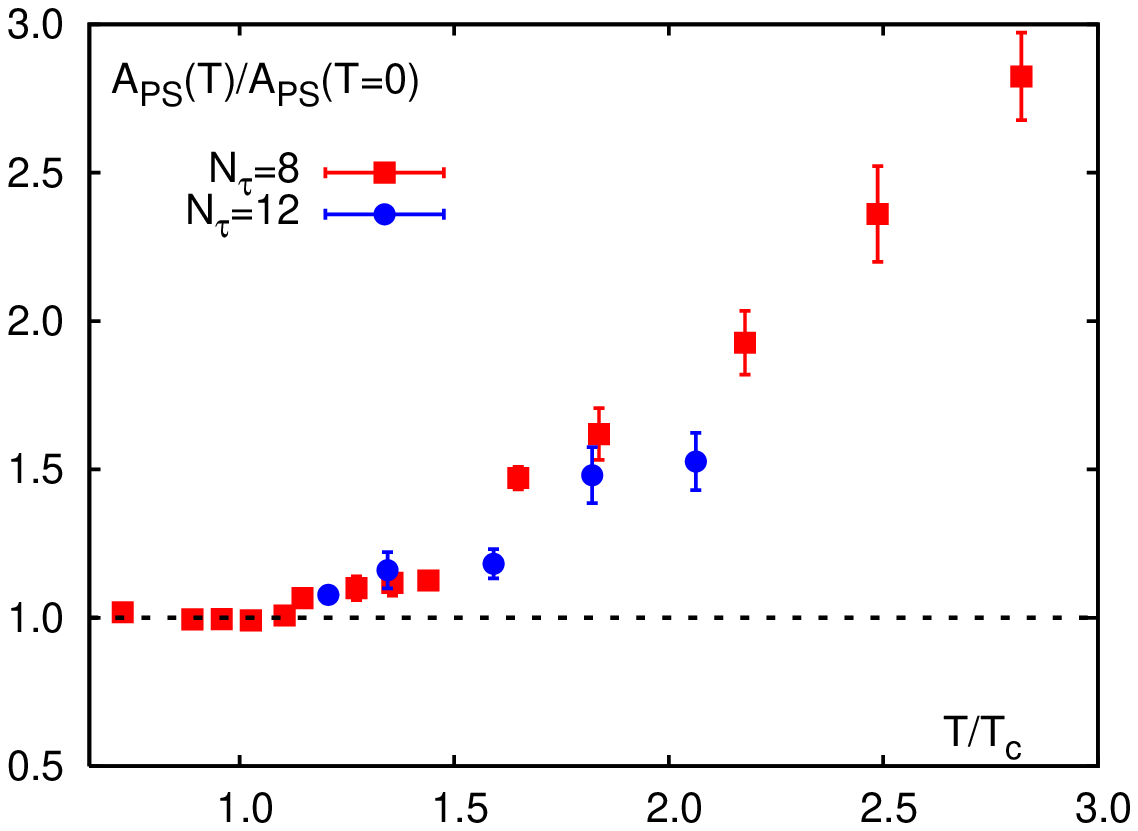} }
\caption{ Pseudo-scalar screening mass (a) and amplitude (b) divided
by the corresponding zero temperature values as function of the temperature.
Filled symbols show the results obtained with point sources, while open symbols
represent the results from the wall sources.  The quenched results are taken from
Ref. \cite{datta04} and correspond to the smallest lattice spacing ($a^{-1}\simeq
9.7$~GeV) used in that study.  (This corresponded to lattices with temporal extent
$N_{\tau}=12,~16$, and $24$). The solid black line is the free quark result (see
text for details).}
\label{fig:ps}
\end{center}
\end{figure}

In Fig.~\ref{fig:ps} we show the screening masses and the amplitudes
divided by the corresponding zero temperature values as function of temperature in
units of the transition temperature $T_c$.  As one can see from the figure the cutoff
effects for the screening masses and the corresponding amplitude are small. At low
temperatures both the masses and the amplitudes agree with their $T=0$ values within
errors.  Above the transition temperature we see small but statistically significant
deviations from the zero temperature result.  We also compare our numerical results
for the screening masses for $T> 1.2T_c$ with the free theory result $2 \sqrt{(\pi
T)^2+m_c^2}$. Here we used $m_c=1.42$~GeV, as will be justified later.

For $T>1.5T_c$ deviations from the zero temperature limit become large and the
screening masses are compatible with the behavior characteristic for unbound $c\bar
c$ pairs. The behavior of the charmonium screening masses is very similar to the
findings of a study in quenched QCD \cite{datta04}. The change in the
temperature dependence of the charmonium screening masses around $T=1.5T_c$ is
similar to the change of the screening masses in the light quark sector around the
QCD transition temperature \cite{our_mscr}. Therefore the change in the behavior of
the charmonium screening masses around $T=1.5T_c$ is likely due to the melting of the
meson states. 

\begin{figure}
\begin{center}
\includegraphics[scale=0.75]{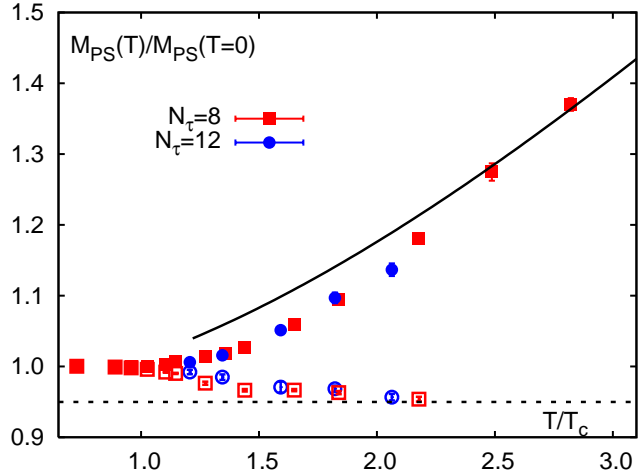}
\caption{Pseudo-scalar charmonium screening masses calculated with anti-periodic
(filled symbols) and periodic (open symbols) boundary conditions as function of the
temperature. Solid (dashed) lines correspond to the free theory prediction with
$m_c=1.42$~GeV for anti-periodic (periodic) boundary conditions.}
\label{fig:pbc}
\end{center}
\end{figure}

As stated above, the dependence of the screening masses on the temporal boundary
conditions can provide additional information about the existence of bound states in
the deconfined medium. Therefore, we also calculated the pseudo-scalar charmonium
screening masses using periodic boundary conditions in the time direction. The
comparison of the screening masses calculated with periodic and anti-periodic
boundary conditions is shown in Fig.~\ref{fig:pbc}.  As one can see from the figure
there is no dependence on the boundary conditions in the low temperature phase of
QCD.  This is consistent with the fact that $c \bar c$ pairs form bound states.  For
$T_c <T< 1.5T_c$ we start seeing small differences in the screening masses calculated
with periodic and anti-periodic boundary conditions.  This might be an indication of
broadening of the $1S$ charmonium state.  At temperatures $T>1.5T_c$ we already see a
strong dependence on the boundary conditions indicating that the fermionic
sub-structure of $c \bar c$ pairs becomes relevant and the charm quark and anti-quark
start becoming sensitive to the lowest Matsubara mode, which is non-zero in the case
of anti-periodic boundary conditions. This is expected to happen when the bound state
is dissolved.  If periodic boundary conditions are used charmonium screening masses
are equal to $2 m_c$ in the limit of very high temperature. If we choose
$m_c=1.42$~GeV at the highest temperature we recover this expectation. Therefore, we
used the value $m_c=1.42$~GeV when comparing with the free field theory prediction.
Note, however, that for $T>2T_c$ the dependence of the screening masses on the
boundary condition is compatible with the free theory expectation regardless what
value of $m_c$ is chosen; {\sl i.e.} we can convert between the results obtained with
periodic boundary condition to the ones obtained with anti-periodic boundary
condition by simply adding the lowest Matsubara frequency in quadrature. 

Let us comment on the behavior of the screening masses in other channels,
{\sl i.e.}  for other choices of $\tilde \phi(X)$. The choice $\tilde
\phi(X)=-(-1)^{y+t}$ (or $\tilde \phi(X)=-(-1)^{x+t}$) corresponds to $\Gamma=\gamma_1$
(or $\Gamma=\gamma_2$) in the non-oscillatory and $\Gamma=\gamma_2\gamma_0$ (or
$\Gamma=\gamma_1\gamma_0$) in the oscillatory part, which in turn correspond to
$J/\psi$ and $h_c$, respectively. We have found that the correlator for this meson
channel is dominated by the $J/\psi$ and we were unable to obtain any signal for the
$h_c$. The qualitative behavior of the $J/\psi$ screening masses was found to be
quite similar to that of the $\eta_c$ \cite{Muk1,Muk2}.
Finally, for
$\Gamma=1$ a pseudo-scalar ($\eta_c$) and a scalar
($\chi_{c0}$) states contribute to the non-oscillatory and oscillatory parts of the
correlator. Similarly the choice $\tilde \phi(X)=-(-1)^y$ (or $\tilde
\phi(X)=-(-1)^x$ ) corresponds to $\gamma_2 \gamma_3$ (or $\gamma_1 \gamma_3$) and
$\gamma_2 \gamma_5$ (or  $\gamma_1 \gamma_5$).  Thus vector ($J/\psi$) and
axial-vector ($\chi_{c1}$) states correspond to the non-oscillatory and oscillatory
parts of the correlator.  Contrary to the situation with light quarks, all three channels
remain to receive both, non-oscillatory and oscillatory contributions above $T_c$.
In fact, up to temperatures of about $2 T_c$ the non-oscillatory part is larger. Thus
the screening masses corresponding to the non-oscillatory part of the correlators can
be reliably calculated and show a temperature dependence that is very similar to that
of the pseudo-scalar ($\eta_c$) and vector ($J/\psi$) screening masses discussed above.  On the other hand, the scalar
and axial-vector screening masses  could be reliably determined only at temperatures
$T>2T_c$. At these temperatures they are consistent with the free theory value.

\section{Conclusions and outlook}

In this paper we have demonstrated that spatial charmonium correlators are useful
tools for studying in-medium charmonium properties and their possible dissolution in
the deconfined phase. The pseudo-scalar screening masses become temperature dependent
just above the crossover temperature of $2+1$ flavor QCD indicating  that medium
modifications of the charmonium spectral function set in gradually.  At temperatures
$T>1.5T_c$ we see already large modifications of the charmonium correlators, and the
temperature dependence of the screening masses is qualitatively the same as  the
behavior expected for unbound $c \bar c$ pairs. At $T>2T_c$ the value of the
screening mass and its dependence on the boundary conditions is in agreement with the
free field limit indicating the absence of charmonium states at these temperatures.
The lattice results presented here are consistent with the expectation based on
potential model and can be used to further constrain potential model calculations of
the charmonium spectral functions. This, of course, will require that those
calculations are extended to finite spatial momentum. 

The investigations presented in this paper can be extended in different ways. While
above the transition temperature cutoff effects are under control, this is not the
case in the hadronic phase. To reduce the cutoff effects at low temperatures
calculations using the Highly Improved Staggered Quark (HISQ)
\cite{Follana:2006rc} action will be necessary. The HISQ action can also reduce
cutoff effects in the light quark sector. It would be furthermore interesting to do
the calculations with a Wilson type action for heavy quarks, like the Fermilab action
\cite{kronfeld}.  Since in this case scalar and axial-vector do not have to be
disentangled from S-wave states this would presumably allow for a more accurate study
of the P-wave charmonia.

\begin{acknowledgments}
This work has been supported by the U.S. Department of Energy under Contract No.
DE-AC02-98CH10886 and the European Union under grant agreement number 238353.  The
numerical computations have been carried out on the apeNEXT computer at Bielefeld
University, QCDOC computers of the RIKEN-BNL research center and the USQCD
collaboration at BNL and the BlueGene/L at the New York Center for Computational
Sciences (NYCCS). We thank S. Datta for providing the code for the screening mass
calculations for QCDOC as well as for the critical comments on the manuscript.
We also thank G. Aarts for his very useful comments and discussions.
\end{acknowledgments}


\begin{thebibliography}{99}

\bibitem{MS86}
T.~Matsui and H.~Satz,
Phys.\ Lett.\ B {\bf 178}, 416 (1986).

\bibitem{qwg}
  N.~Brambilla {\it et al.} 
  Eur.\ Phys.\ J.\ C {\bf 71}, 1534 (2011).

\bibitem{qgp4}
  A.~Bazavov, P.~Petreczky and A.~Velytsky,
Quark Gluon Plasma 4, p. 61 , World Scientific 2010, Eds. R.C. Hwa, X.-N. Wang
  [arXiv:0904.1748 [hep-ph]]; \\ R. Rapp and H. van Hees, ibid. p. 111.

\bibitem{sps}
  R.~Arnaldi {\it et al.}  (NA60 Collaboration),
  Nucl.\ Phys.\  A {\bf 783}, 261 (2007).

\bibitem{rhic}
R. Granier de Cassagnac, J. Phys. G. {\bf 35}, 104023 (2008).

\bibitem{lhc}
  P.~Pillot (ALICE Coll.),
  arXiv:1108.3795 [hep-ex];\\
  G.~Aad {\it et al.}  (Atlas Coll.),
  Phys.\ Lett.\  B {\bf 697}, 294 (2011);\\
  C.~Silvestre (CMS Coll.),
  arXiv:1108.5077 [hep-ex].

\bibitem{asakawa01}
  M.~Asakawa, T.~Hatsuda and Y.~Nakahara,
  Prog.\ Part.\ Nucl.\ Phys.\  {\bf 46}, 459 (2001).

\bibitem{ineslat01}
  I.~Wetzorke {\it et. al},
  Nucl.\ Phys.\ Proc.\ Suppl.\  {\bf 106}, 510 (2002).


\bibitem{umeda02}
T.~Umeda {\it et al.},
Eur.\ Phys.\ J.  C {\bf 39S1}, 9 (2005).

\bibitem{asakawa03}
M.~Asakawa and T.~Hatsuda,
Phys.\ Rev.\ Lett.\  {\bf 92}, 012001 (2004).

\bibitem{datta04}
S.~Datta {\it et al.},
Phys.\ Rev.\ D {\bf 69}, 094507 (2004). 

   
\bibitem{jako06}
  A.~Jakovac {\it et al.},
  Phys.\ Rev.\  D {\bf 75}, 014506 (2007).


\bibitem{alton07}
  G.~Aarts {\it et al.},
  Phys.\ Rev.\ D {\bf 76}, 094513 (2007).

\bibitem{Iida:2006mv}
  H.~Iida {\it et al.},
  Phys.\ Rev.\ D {\bf 74}, 074502 (2006).


\bibitem{Ohno:2011zc}
  H.~Ohno {\it et al.} (WHOT-QCD Collaboration),
  arXiv:1104.3384 [hep-lat].

\bibitem{umeda07}
  T.~Umeda,
  Phys.\ Rev.\ D {\bf 75}, 094502 (2007).

\bibitem{petr08}
  P.~Petreczky,
  Eur.\ Phys.\ J.\ C {\bf 62}, 85 (2009).

\bibitem{ding10}
  H.~-T.~Ding {\it et al.},
  PoSLATTICE\ {\bf 2010}, 180  (2010).


\bibitem{Aarts:2010ek} 
  G.~Aarts, S.~Kim, M.~P.~Lombardo, M.~B.~Oktay, S.~M.~Ryan, D.~K.~Sinclair and J.~-I.~Skullerud,
  Phys.\ Rev.\ Lett.\  {\bf 106}, 061602 (2011)


\bibitem{f1}
  O.~Kaczmarek {\it et al.},
  Phys.\ Lett.\  B {\bf 543}, 41 (2002); 
  Phys.\ Rev.\  D {\bf 70}, 074505 (2004)
  [Erratum-ibid.\  D {\bf 72}, 059903 (2005)];
  P.~Petreczky and K.~Petrov,
  Phys.\ Rev.\  D {\bf 70}, 054503 (2004);
  O.~Kaczmarek and F.~Zantow,
  Phys.\ Rev.\  D {\bf 71}, 114510 (2005).

\bibitem{digal01}
S.~Digal {\it et al.},
Phys.\ Rev.\ D {\bf 64}, 094015 (2001).

\bibitem{mocsy07}
  A.~Mocsy and P.~Petreczky,
  Phys.\ Rev.\ Lett.\  {\bf 99}, 211602 (2007);
  Phys.\ Rev.\ D {\bf 77}, 014501 (2008);
  Phys.\ Rev.\ D {\bf 73}, 074007  (2006).

\bibitem{blaschke04}
  D.~Blaschke {\it et al.},
  Eur.\ Phys.\ J. C {\bf 43}, 81 (2005).

\bibitem{laine06}
  M.~Laine {\it et. al},
  JHEP {\bf 0703}, 054 (2007).

\bibitem{nora08}
  N.~Brambilla {\it et al.},
  Phys.\ Rev.\ D {\bf 78}, 014017 (2008).


\bibitem{miao10}
  C.~Miao, A.~Mocsy and P.~Petreczky,
  Nucl.\ Phys.\ A {\bf 855}, 125 (2011);\\
  M.~Margotta {\it et al.},
  Phys.\ Rev.\ D {\bf 83}, 105019 (2011).

\bibitem{riek}
  F.~Riek and R.~Rapp,
  New J.\ Phys.\ \ {\bf 13}, 045007  (2011).


\bibitem{boyd94}
  G.~Boyd {\it et al.},
  Z.\ Phys.\ C\ {\bf 64}, 331  (1994).

\bibitem{Muk1} 
  S.~Mukherjee,
  Nucl.\ Phys.\ A\ {\bf 820}, 283C  (2009).

\bibitem{Muk2} 
  S.~Mukherjee,
  PoSCONFINEMENT\ {\bf 2008}, 116  (2008).


\bibitem{our_mscr}
  M.~Cheng {\it et al.},
  Eur.\ Phys.\ J. C {\bf 71}, 1564 (2011).


\bibitem{p4-1} 
  U.~M.~Heller, F.~Karsch and B.~Sturm,
  Phys.\ Rev.\ D\ {\bf 60}, 114502  (1999).

\bibitem{p4}
  F.~Karsch, E.~Laermann and A.~Peikert,
  Nucl.\ Phys.\ B\ {\bf 605}, 579  (2001).

\bibitem{rbcbi07}
  M.~Cheng {\it et al.},
  Phys.\ Rev.\ D {\bf 77}, 014511 (2008).
\bibitem{hoteos}

  A.~Bazavov {\it et al.},
  Phys.\ Rev.\ D {\bf 80}, 014504 (2009).

\bibitem{gray05}
  A.~Gray {\it et al.},
  Phys.\ Rev.\  D {\bf 72}, 094507 (2005).

\bibitem{sommer}
  R.~Sommer,
  Nucl.\ Phys.\ B {\bf 411}, 839 (1994).

\bibitem{rbcbi06}
  M.~Cheng {\it et al.},
  Phys.\ Rev.\ D\ {\bf 74}, 054507  (2006).

\bibitem{hotqcd2}
  A.~Bazavov {\it et al.},
  Phys.\ Rev.\ D\ {bf 85}, 054503 (2012).

\bibitem{cheng-1} 
  M.~Cheng,
  PoSLATTICE {\bf 2007}, 173  (2007).

\bibitem{cheng-2} 
  M.~Cheng, Ph.D Thesis,
  ``The QCD equation of state with charm quarks from lattice QCD,''
  (2008), AAT-3333321.

\bibitem{kluberg}
H. Kluberg-Stern {\it et al.}, Nucl. Phys. B {\bf 220}, 447 (1983).

\bibitem{golterman}
M. F. L. Golterman, Nucl. Phys. B {\bf 273}, 663 (1986).

\bibitem{altmeyer}
R. Altmeyer {\it et al.}, Nucl. Phys. B {\bf 389}, 445 (1993).

\bibitem{florkowski}
  W.~Florkowski and B.~L.~Friman,
  Z.\ Phys.\ A {\bf 347}, 271 (1994).


\bibitem{Follana:2006rc} 
E. Follana {\it et al.}\ [HPQCD and UKQCD Collaborations], 
Phys.\ Rev.\ D\ {\bf 75}, 054502 (2007).

\bibitem{kronfeld}
  A.~X.~El-Khadra, A.~S.~Kronfeld and P.~B.~Mackenzie,
  Phys.\ Rev.\ D\ {\bf 55}, 3933  (1997).


\end{thebibliography}
\end{document}